\documentclass[twocolumn,showpacs,preprintnumbers,superscriptaddress,floatfix,nofootinbib]{revtex4}
\usepackage{amssymb}
\usepackage{amsmath}
\usepackage{graphicx}
\usepackage{dcolumn}
\usepackage{bm}
\usepackage[subfigure]{graphfig}
\usepackage{makecell}

\begin{document}

\title{Discovery potential for the neutral charmoniumlike $Z^{0}(4200)$ by $%
\bar{p}p$ annihilation}
\author{Xiao-Yun Wang}
\thanks{Corresponding author: xywang@impcas.ac.cn}
\affiliation{Institute of Modern Physics, Chinese Academy of Sciences, Lanzhou 730000,
China}
\affiliation{University of Chinese Academy of Sciences, Beijing 100049, China}
\affiliation{Research Center for Hadron and CSR Physics, Institute of Modern Physics of
CAS and Lanzhou University, Lanzhou 730000, China}
\author{Xu-Rong Chen}
\affiliation{Institute of Modern Physics, Chinese Academy of Sciences, Lanzhou 730000,
China}
\affiliation{Research Center for Hadron and CSR Physics, Institute of Modern Physics of
CAS and Lanzhou University, Lanzhou 730000, China}

\begin{abstract}
Inspired by the observation of charmoniumlike $Z(4200)$, we explore the
discovery potential of the neutral $Z^{0}(4200)$ production by
antiproton-proton annihilation with an effective Lagrangian approach. By
investigating the $\bar{p}p\rightarrow J/\psi \pi ^{0}$ process including
the $Z^{0}(4200)$ signal and background contributions, it is found that the
center of mass energy $E_{c.m.}\simeq 4.0-4.5$ GeV is the best energy window
for searching the neutral $Z^{0}(4200)$, where the signal can be clearly
distinguished from background. The relevant calculations is not only helpful
to search for the neutral $Z^{0}(4200)$ in the future experiment, but also
will promote the understanding the nature and production mechanism of
neutral $Z^{0}(4200)$ better.
\end{abstract}

\pacs{14.40.Lb, 13.75.Cs,11.10.Ef}
\maketitle

\section{Introduction}

In the conventional constituent quark model (CQM), all mesons can be
described as quark-antiquark ($q\bar{q}$) states. Such as the charmonium
state $J/\psi $, it is composed of a $c\bar{c}$ quark pair. However, a
series of charmoniumlike states (including $X(3872)$, $Y(4260)$, $Y(4008)$, $%
Z(3900)$, $Z(4025)$, $Z(4020)$, $Z(4050)$, $Z(4250)$, $Z(4430)$ and $Z(4200)$
etc.) have been observed in experiments \cite%
{liux14,cms13,belle100,lhc14,belle14}, many of which do not fit into the $q%
\bar{q}$ meson spectrum in the classical CQM. In particular, some charged
charmonium-like $Z$ states are considered to have a minimal quark content of
$\left\vert c\bar{c}u\bar{d}\right\rangle $ ($Z^{+}$) or $\left\vert c\bar{c}%
\bar{u}d\right\rangle $ ($Z^{-}$), which have attracted great interest in
experimental and theoretical aspect \cite%
{liux14,cms13,belle100,lhc14,belle14,tma15,zl14,wei11,wei12,wei15,wzg15}.
These charmoniumlike $Z$ states are interpreted as a tetraquark, a hadron
molecule or just a cusp effect, and so on \cite%
{liux14,zl14,wei11,wei12,wei15,wzg15}. In addition, a series of hidden charm
baryons also have been investigated in Refs. \cite{wjj10,wjj11,xyw15}, which
were considered as the pentaquark candidates and composed of $\left\vert c%
\bar{c}qqq\right\rangle $. On the one hand, these studies enriched the
picture of exotic states. On the other hand, it should be noted that our
understanding of these exotic states is far from being sufficient. Thus,
more theoretical and experimental works are needed.

Recently, a new charged charmonium-like $Z^{+}(4200)$ decaying to $J/\psi
\pi ^{+}$ has been observed by the Belle Collaboration with a significance
of 6.2$\sigma $ \cite{belle14}. The relevant experiment results show that
the spin-parity quantum number of $Z^{+}(4200)$ favors $1^{+}$. Furthermore,
its mass and width are measured to be \cite{belle14}
\begin{eqnarray*}
M &=&4196_{-29-13}^{+31+17}\text{ MeV/}c^{2}, \\
\Gamma &=&370_{-70-132}^{+70+70}\text{ MeV.}
\end{eqnarray*}%
It is notice that the $Z^{+}(4200)$ has the largest width among those
charmonium-like $Z$ states which have been observed in experiment. In Ref.
\cite{zl14}, by analyzing all the available experimental information about
charmonium-like states within the framework of the color-magnetic
interaction, the $Z(4200)$ was supposed to be a very promising candidate of
the lowest axial-vector hidden-charm tetraquark state and its dominant decay
should be $J/\psi \pi $. In Ref. \cite{wei11,wei12,wei15}, the relevant
results also support the tetraquark interpretation of $Z(4200)$ in the frame
of QCD sum rule. Moreover, with QCD sum rule approach, the $Z(4200)$ was
described as the molecule-like state in Ref. \cite{wzg15} These experimental
and theoretical results indicate that the $Z(4200)$ is an ideal candidate
for investigating and understanding the nature of exotic charmonium-like
states.

As of now, the charmonium-like states are only observed in four ways
including the $\gamma \gamma $ fusion process, $e^{+}e^{-}$ annihilation, $B$
meson decay and hidden-charm dipion decays of higher charmonia or
charmonium-like states. Obviously, it is an important topic to study the the
production of the charmonium-like state in different processes. As mentioned
above, $Z(4200)$ was only observed in the $B$ meson decay. It is natural to
ask whether $Z(4200)$ can be found in other processes.

In addition, we note that searching for the neutral partner of the
charmonium-like or bottomoniumlike state also aroused the great interest
both in experiment and theory. The first observation of neutral $Z_{c}(3900)$
has been reported by CLEO-c \cite{tx13}. Later, the neutral bottomoniumlike $%
Z_{b}^{0}(10610)$ and charmonium-like $Z_{c}^{0}(4020)$ have been observed
by Belle and BESIII \cite{zb13,zc14}, respectively. In theory, the
production of the neutral $Z^{0}(4430)$ in $\bar{p}p\rightarrow \psi
^{\prime }(2s)\pi ^{0}$ reaction was investigated in our previous work \cite%
{wang15}. These studies not only help in confirming these charmonium-like
state, but also opens a window to investigate the nature and production
mechanism of exotic state beyond the conventional $q\bar{q}$ states.

Since the $Z(4200)$ was observed in $J/\psi \pi $ channel \cite{belle14} and
its dominant decay mode is very likely to be $J/\psi \pi $ \cite{zl14}, the
neutral $Z^{0}(4200)$ should has a coupling with $J/\psi \pi ^{0}$. Besides,
the tetraquark or molecule-like state can be regard as a general four-quark
state \cite{wei15}, which means that the neutral $Z^{0}(4200)$ probably is
composed by $\left\vert c\bar{c}u\bar{u}\right\rangle $ or $\left\vert c\bar{%
c}d\bar{d}\right\rangle $. According to the OZI rule \cite{ozi}, one can
speculate that the partial decay width of $Z^{0}(4200)\rightarrow \bar{p}p$
may be larger than that of $J/\psi \rightarrow \bar{p}p$. Therefore, the $%
\bar{p}p\rightarrow J/\psi \pi ^{0}$ reaction is probably an ideal channel
for searching and studying the neutral $Z^{0}(4200)$.

In this work, with an effective Lagrangian approach, the production of
neutral $Z^{0}(4200)$ in $\bar{p}p\rightarrow J/\psi \pi ^{0}$ reaction are
investigated for the first time. Furthermore, in light of the situations of
PANDA detector at FAIR@GSI \cite{panda,uw11,cs12}, the feasibility of
searching the neutral $Z^{0}(4200)$ by $\bar{p}p$ annihilation is discussed,
which can provide valuable information to future experimental exploration of
neutral $Z^{0}(4200)$.

This paper is organized as follows. After an introduction, we present the
investigate method and formalism for $Z^{0}(4200)$ production. In Sec. III,
the background contributions to the $J/\psi \pi ^{0}$ final states are
discussed. The numerical result are given in Sec. IV. Finally, This paper
ends with the discussion and conclusion.

\section{The $Z^{0}(4200)$ yield by the antiproton-proton scattering}

It will be difficult to study $Z^{0}(4200)$ at quark-gluon level in the now
energy range. Therefore, the effective Lagrangian method in terms of hadrons
will be used in our research.

\subsection{Feynman diagrams and effective interaction Lagrangian densities}

The basic tree level Feynman diagram for the production of $Z^{0}(4200)$ in $%
\bar{p}p\rightarrow J/\psi \pi ^{0}$ reaction through $s$-channel is
depicted in Fig. 1.
\begin{figure}[bp]
\centering
\includegraphics[scale=0.6]{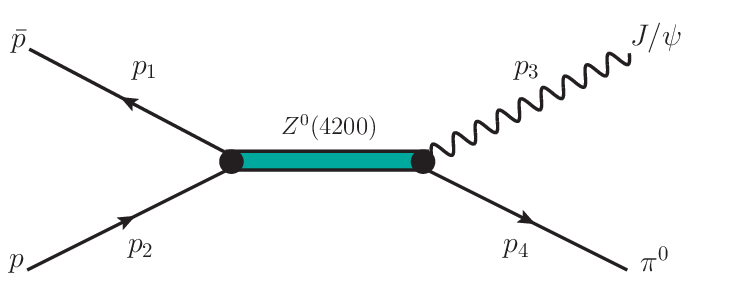}
\caption{(Color online) The production of $Z^{0}(4200)$ through $p\bar{p}$
collision.}
\end{figure}
For the $Z(4200)$, its quantum number of spin-parity has been determined by
Belle Collaboration to be $J^{P}=1^{+}$ \cite{belle14}. Therefore, the
relevant effective Lagrangian for the vertices of $Z\bar{p}p$ and $Z\psi \pi
$\footnote{%
For simplicity, we use $Z$ and $\psi $ denote $Z^{0}(4200)$ and $J\psi $,
respectively.} read as \cite{gy08},

\begin{eqnarray}
\mathcal{L}_{Z\bar{p}p} &=&g_{Z\bar{p}p}\bar{\phi}_{\bar{p}}\gamma ^{\mu
}\gamma _{5}\phi _{p}Z_{\mu }, \\
\mathcal{L}_{Z\psi \pi } &=&\frac{g_{Z\psi \pi }}{M_{Z}}(\partial ^{\mu
}\psi ^{\nu }\partial _{\mu }\pi Z_{\nu }-\partial ^{\mu }\psi ^{\nu
}\partial _{\nu }\pi Z_{\mu }),
\end{eqnarray}%
where $Z$, $\psi $ and $\phi $ denote the fields of $Z(4200)$, $J/\psi $ and
nucleon, respectively. The $g_{Z\bar{p}p}$ and $g_{Z\psi \pi }$ are the
coupling constants. Considering the size of the hadrons, we introduce the
general form factor for the intermediate $Z^{0}(4200)$ as used in Refs. \cite%
{mosel98,mosel99,vs05},%
\begin{equation}
\mathcal{F}_{Z}(q^{2})=\frac{\Lambda _{Z}^{4}}{\Lambda
_{Z}^{4}+(q^{2}-M_{Z}^{2})^{2}},
\end{equation}%
where $q$, $M_{Z}$, and $\Lambda _{Z}$ are the 4-momentum, mass, and cut-off
parameters for the intermediate $Z^{0}(4200)$, respectively.

\subsection{Coupling constants and the OZI analysis in the process of $%
Z^{0}(4200)\rightarrow \bar{p}p$}

\begin{figure}[tbp]
\begin{center}
\includegraphics[scale=0.6]{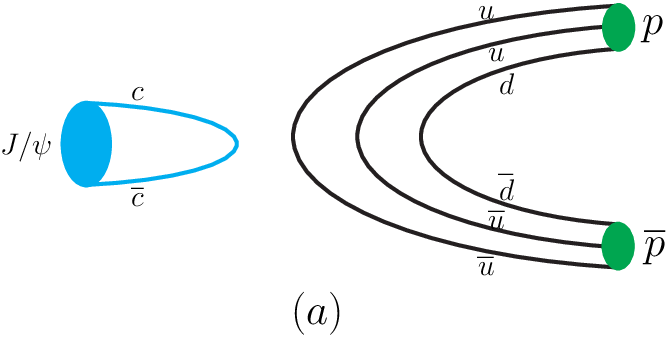} \hspace{80pt} %
\includegraphics[scale=0.6]{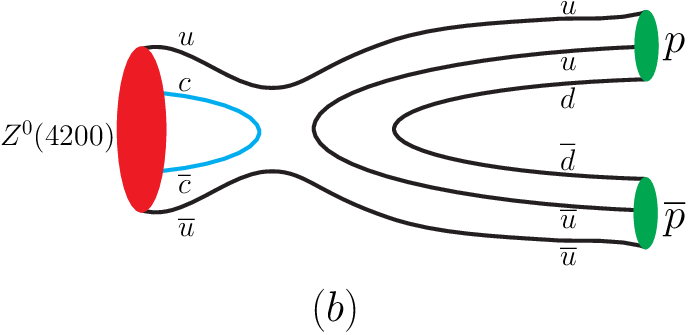}
\end{center}
\caption{(Color online) The quark level diagram depicting $J/\protect\psi %
\to \bar{p}p$ decay [(a)] and $Z^{0}(4200) \to \bar{p}p$ decay [(b)].}
\label{ozi}
\end{figure}

With the effective Lagrangians above, the coupling constant $g_{Z\bar{p}p}$
and $g_{Z\psi \pi }$ can be determined by the partial decay widths $\Gamma
_{Z^{0}(4200)\rightarrow \bar{p}p}$ and $\Gamma _{Z^{0}(4200)\rightarrow
J/\psi \pi ^{0}}$, respectively,

\begin{equation}
\Gamma _{Z^{0}(4200)\rightarrow \bar{p}p}=\left( g_{Z\bar{p}p}\right) ^{2}%
\frac{2}{3\pi M_{Z}^{2}}|\vec{p}_{N}^{~\mathrm{c.m.}}|^{3},
\end{equation}%
\begin{eqnarray}
\Gamma _{Z^{0}(4200)\rightarrow J/\psi \pi ^{0}} =\left( \frac{g_{Z\psi \pi }%
}{M_{Z}}\right) ^{2}\frac{|\vec{p}_{\pi }^{~\mathrm{c.m.}}|}{24\pi M_{Z}^{2}}
\notag \\
\times \left[ \frac{(M_{Z}^{2}-m_{\psi }^{2}-m_{\pi }^{2})^{2}}{2}+m_{\psi
}^{2}E_{\pi }^{2}\right] ,
\end{eqnarray}%
with

\begin{eqnarray}
|\vec{p}_{N}^{~\mathrm{c.m.}}| &=&\frac{\lambda ^{1/2}(M_{Z}^{2},m_{\bar{p}%
}^{2},m_{p}^{2})}{2M_{Z}}, \\
|\vec{p}_{\pi }^{~\mathrm{c.m.}}| &=&\frac{\lambda ^{1/2}(M_{Z}^{2},m_{\psi
}^{2},m_{\pi }^{2})}{2M_{Z}}, \\
E_{\pi } &=&\sqrt{|\vec{p}_{\pi }^{~\mathrm{c.m.}}|^{2}+m_{\pi }^{2}},
\end{eqnarray}%
where $\lambda $ is the K$\ddot{a}$llen function with $\lambda
(x,y,z)=(x-y-z)^{2}-4yz$.

As of now, no relevant experiment datas about $\Gamma _{Z(4200)\rightarrow
\bar{p}p}$ and $\Gamma _{Z(4200)\rightarrow J/\psi \pi }$ are available \cite%
{pdg14}. However, in Refs. \cite{wei15,wzg15} the decay width $\Gamma
_{Z(4200)\rightarrow J/\psi \pi }=87.3\pm 47.1$ MeV or $\Gamma
_{Z_{c}(4200)\rightarrow J/\psi \pi }=24.6$ MeV were obtained with QCD sum
rule by assuming that the $Z(4200)$ is a tetraquark state or molecule-like
state, respectively. Thus we get the coupling constants $g_{Z\psi \pi
}/M_{z}=1.73,0.918$ GeV$^{-1}$, which correspond to the partial decay width $%
\Gamma _{Z(4200)\rightarrow J/\psi \pi }=87.3,24.6$ MeV.

For the partial decay width of $Z^{0}(4200)\rightarrow \bar{p}p$, we try to
obtain it by analyzing and comparing with the OZI suppressed process of $%
J/\psi \rightarrow \bar{p}p$. According to constituent quark model, the
traditional charmonium $J/\psi $ is regarded as a pure $c\overline{c}$
state, while there are only up and down quarks (antiquarks) in the proton
(antiproton). Thus the $J/\psi \rightarrow \bar{p}p$ decay is actually a
disconnected process and at least need three gluons to connect it (as shown
in Fig. 2(a)). In the frame of the Okubo-Zweig-Iizuka (OZI) rule \cite{ozi},
the incidence of $J/\psi \rightarrow \bar{p}p$ process is greatly
suppressed. With the total decay width and branch ratios of $J/\psi $ listed
in PDG \cite{pdg14}, one get the partial decay width $\Gamma _{J/\psi
\rightarrow \bar{p}p}\simeq 0.2$ keV, which is indeed a small value and
consistent with the prediction by OZI rule \cite{ozi}. Since the neutral $%
Z^{0}(4200)$ may be have minimal quark content of ($c\overline{c}u\overline{u%
}$) or ($c\overline{c}d\overline{d}$), as seen in Fig. 2(b), the $%
Z^{0}(4200)\rightarrow \bar{p}p$ reaction is a connected whole and it is an
OZI allowed process \cite{ozi}. Therefore, in principle, the probability of $%
Z^{0}(4200)$ decay to $\bar{p}p$ should be higher than $J/\psi \rightarrow
\bar{p}p,$ which may be part of reason that the total decay width of $%
Z(4200) $ is 3 order larger than the total width of $J/\psi .$ Accordingly
we speculate that the partial width $\Gamma _{Z^{0}(4200)\rightarrow \bar{p}%
p}$ may well be at least three magnitude larger than $\Gamma _{J/\psi
\rightarrow \bar{p}p}$. Then we get $g_{Z\bar{p}p}\simeq 0.05$ by taking $%
\Gamma _{Z^{0}(4200)\rightarrow \bar{p}p}=200$ keV.

\subsection{Amplitude}

Following the Feynman rules and using above Lagrangian densities, we can
obtain the invariant amplitude $\mathcal{M}_{Z}^{signal}$ for the $\bar{p}%
(p_{1})p(p_{2})\rightarrow J/\psi (p_{3})\pi ^{0}(p_{4})$ reaction through $%
s $-channel as shown in Fig. 1,%
\begin{eqnarray}
\mathcal{M}_{Z}^{signal} &\mathcal{=}&\frac{g_{Z\bar{p}p}g_{Z\psi \pi }%
\mathcal{F}_{Z}(q^{2})}{M_{z}}\varepsilon ^{\nu }(p_{3})\cdot (p_{3}\cdot
p_{4}g_{\mu \nu }-p_{3\mu }p_{4\nu })  \notag \\
&&\times G_{Z}^{\mu \alpha }(q)\overline{v}(p_{1})\gamma _{\alpha }\gamma
_{5}u(p_{2}),
\end{eqnarray}%
where $G_{Z}^{\mu \alpha }$ are the propagators of the $Z^{0}(4200),$ taking
the Breit-Wigner form \cite{wh02},%
\begin{equation}
G_{Z}^{\mu \alpha }(q)=i\frac{\mathcal{P}^{(1)}}{q^{2}-M_{Z}^{2}+iM_{Z}%
\Gamma _{Z}}.
\end{equation}%
Here
\begin{eqnarray}
\mathcal{P}^{(1)} &=&\sum_{spins}\varepsilon ^{\mu }\cdot \varepsilon
_{\alpha }^{\ast }=\tilde{g}^{\mu \alpha }(q)  \notag \\
&=&-g^{\mu \alpha }+q^{\mu }q^{\alpha }/q^{2}
\end{eqnarray}%
is the projection operator for the state with spin-1.

\section{\protect\bigskip The background analysis and cross section}

\begin{figure}[tbp]
\centering
\includegraphics[height=2.8cm,width=0.48\textwidth]{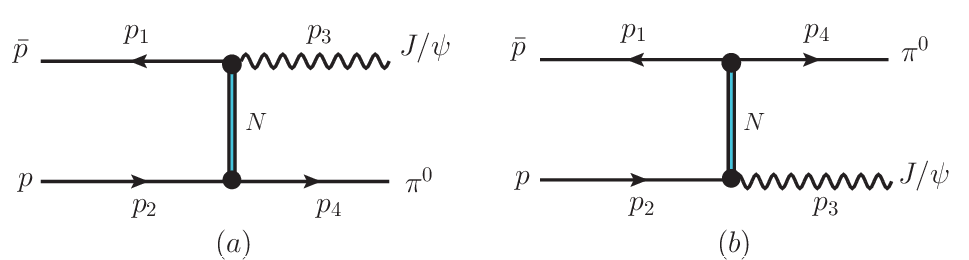}
\caption{(Color online) Feynman diagrams for the nucleon exchange in $%
\overline{p}p\rightarrow J/\protect\psi \protect\pi ^{0}$ reaction.}
\end{figure}

Fig. 3 show the $\bar{p}(p_{1})p(p_{2})\rightarrow J/\psi (p_{3})\pi
^{0}(p_{4})$ process through t-channel (a) and u-channel (b) by exchanging a
proton, which can be regarded as the main background contributions for the
production of $Z^{0}(4200)$ as described in Fig. 1.

The Lagrangian densities for the vertices of $\bar{p}p\pi ^{0}$ and $J/\psi
\bar{p}p$ read as \cite{cd06}%
\begin{eqnarray}
\mathcal{L}_{\bar{p}p\pi ^{0}} &=&-ig_{\bar{p}p\pi }\bar{\phi}_{\bar{p}%
}\gamma _{5}\tau \cdot \pi \phi _{p}, \\
\mathcal{L}_{\bar{p}p\psi } &=&-g_{\bar{p}p\psi }\bar{\phi}_{\bar{p}}\gamma
^{\mu }\phi _{p}\psi _{\mu },
\end{eqnarray}%
where $\psi $ and $\phi $ denote the fields of $J/\psi $ and nucleon,
respectively, while $\tau $ is Pauli matrix.

The coupling constant $g_{\bar{p}p\pi }=13.5$ is adopted \cite{zwl00}, while
coupling constant $g_{\bar{p}p\psi }$ is determined by partial decay widths:

\begin{equation}
\Gamma _{\psi \rightarrow \bar{p}p}=\left( g_{\bar{p}p\psi }\right) ^{2}%
\frac{\left\vert \vec{p}_{c.m.}\right\vert }{6\pi m_{\psi }^{2}}\left(
m_{\psi }^{2}+2m_{p}^{2}\right)
\end{equation}%
with%
\begin{equation}
\vec{p}_{c.m.}=\frac{\sqrt{m_{\psi }^{2}-4m_{p}^{2}}}{2}.
\end{equation}%
Thus we get $g_{\bar{p}p\psi }\simeq 1.6\times 10^{-3}$, which is calculated
by the measured branching fractions and total widths of $J/\psi $ ($m_{\psi
}=3096.916$ MeV and $\Gamma _{\psi }=92.9$ keV) \cite{pdg14}.

The monopole form factors for the $\bar{p}p\pi ^{0}$ and $J/\psi \bar{p}p$
vertices are introduced as the same as Bonn potential model \cite{rm87}:%
\begin{equation}
\mathcal{F(}q_{i}^{2}\mathcal{)=}\frac{\Lambda _{N}^{2}-m_{p}^{2}}{\Lambda
_{N}^{2}-q_{i}^{2}},\text{ }i=t,u
\end{equation}%
where $\Lambda _{N}$, $m_{p}$ and $q_{i}$ ($q_{t}=(p_{3}-p_{1})$ and $%
q_{u}=(p_{4}-p_{1})$) are the cut-off parameter, mass and four-momentum of
the exchanged proton, respectively.

According to the Feynman rules and above equations, the full invariant
amplitude $\mathcal{M}_{N}=\mathcal{M}_{N}^{t}+\mathcal{M}_{N}^{u}$ for the
background as depicted in Fig. 3 can be obtained,

\begin{eqnarray}
\mathcal{M}_{N} &=&\mathcal{M}_{N}^{t}+\mathcal{M}_{N}^{u}  \notag \\
&\mathcal{=}&g_{\bar{p}p\pi }g_{\bar{p}p\psi }\overline{v}_{\bar{p}}(p_{1})%
\Big\{\gamma _{\mu }\varepsilon ^{\mu }(p_{3})\frac{\rlap{$\slash$}%
q_{t}+m_{p}}{q_{t}^{2}-m_{p}^{2}}\gamma ^{5}\mathcal{F}^{2}\mathcal{(}%
q_{t}^{2}\mathcal{)}  \notag \\
&&+\gamma ^{5}\frac{\rlap{$\slash$}q_{u}+m_{p}}{q_{u}^{2}-m_{p}^{2}}\gamma
_{\mu }\varepsilon ^{\mu }(p_{3})\mathcal{F}^{2}\mathcal{(}q_{u}^{2}\mathcal{%
)}\Big\}u_{p}(p_{2}).
\end{eqnarray}%
With the amplitudes listed in Eqs. (9) and (17), we get the square of the
total invariant amplitude\footnote{%
In principle, the interference between amplitudes for the signal and the
non-resonant background should be considered. Since we do not now have
experimental data, we take the relative phase between different amplitudes
as zero in the present work. Thus the total cross section for the $\bar{p}%
p\rightarrow J/\psi \pi ^{0}$ process obtained by us is an upper estimate.}%
\begin{equation}
\left\vert \mathcal{M}\right\vert ^{2}=\sum \left\vert \mathcal{M}%
_{Z}^{signal}+\mathcal{M}_{N}\right\vert ^{2}\text{.}
\end{equation}%
We define $s=q^{2}=(p_{1}+p_{2})^{2},$ then the unpolarized differential
cross section for the reaction $\bar{p}(p_{1})p(p_{2})\rightarrow J/\psi
(p_{3})\pi ^{0}(p_{4})$ at the center of mass (c.m.) frame is as follows:%
\begin{equation}
\frac{d\sigma }{d\cos \theta }=\frac{1}{32\pi s}\frac{\left\vert \vec{p}%
_{3}^{~\mathrm{c.m.}}\right\vert }{\left\vert \vec{p}_{1}^{~\mathrm{c.m.}%
}\right\vert }\left( \frac{1}{4}\sum\limits_{spins}\left\vert \mathcal{M}%
\right\vert ^{2}\right) ,
\end{equation}%
where $\vec{p}_{1}^{~\mathrm{c.m.}}$ and $\vec{p}_{3}^{~\mathrm{c.m.}}$ are
the three-momentum of initial anti-proton and final $J/\psi $, while $\theta
$ denotes the angle of the outgoing $J/\psi $ meson relative to the
anti-proton beam direction in the c.m. frame. The total cross section can be
easily obtained by integrating the above equation.

\section{Numerical results and discussion}

\begin{figure}[tbph]
\centering
\includegraphics[scale=0.4]{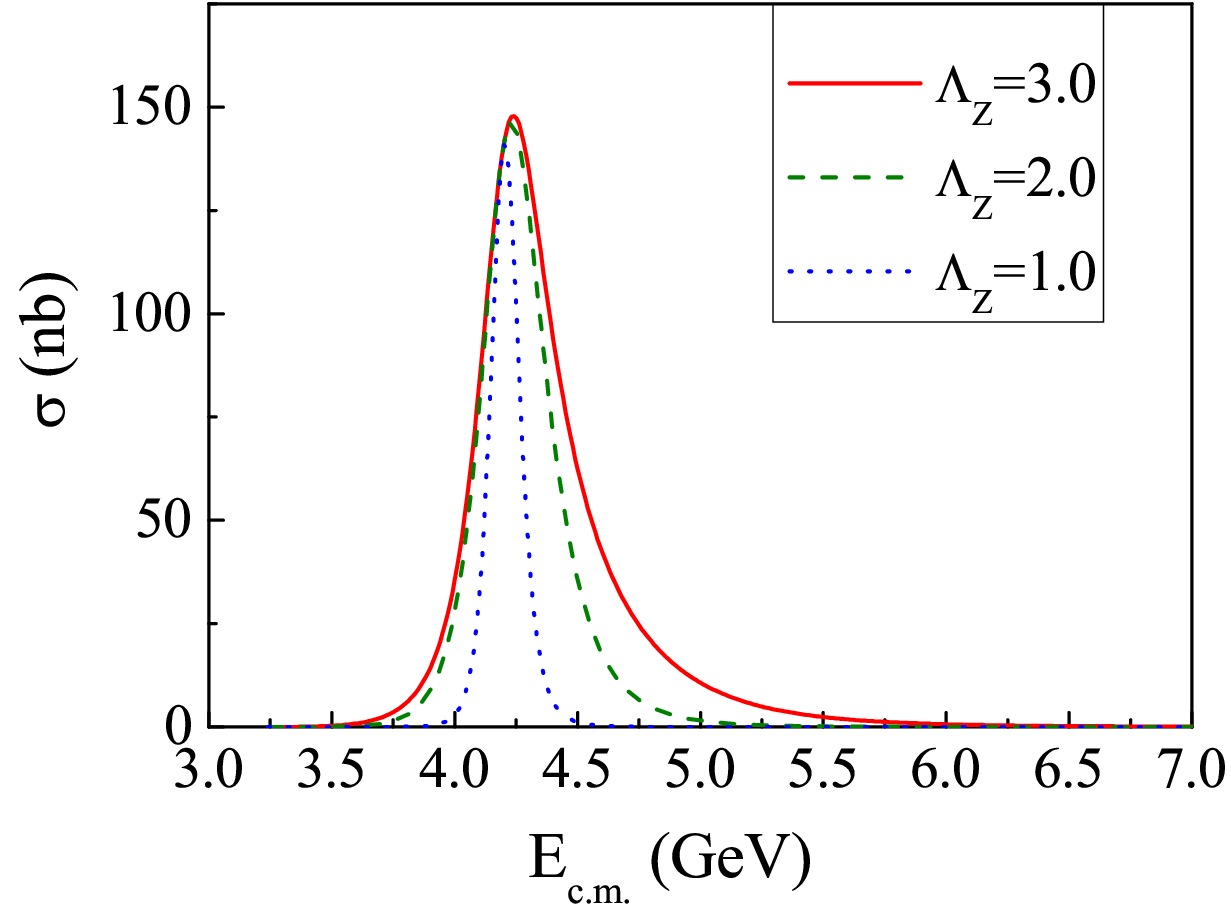}
\caption{(Color online) The energy dependence of cross section for the
production of $Z^{0}(4200)$ through s-channel with the different typical
cut-off $\Lambda _{Z}$. Here, the partial decay width is taken as $\Gamma
_{Z(4200)\rightarrow J/\protect\psi \protect\pi }=87.3$ MeV.}
\end{figure}

\begin{figure}[bph]
\centering
\includegraphics[scale=0.4]{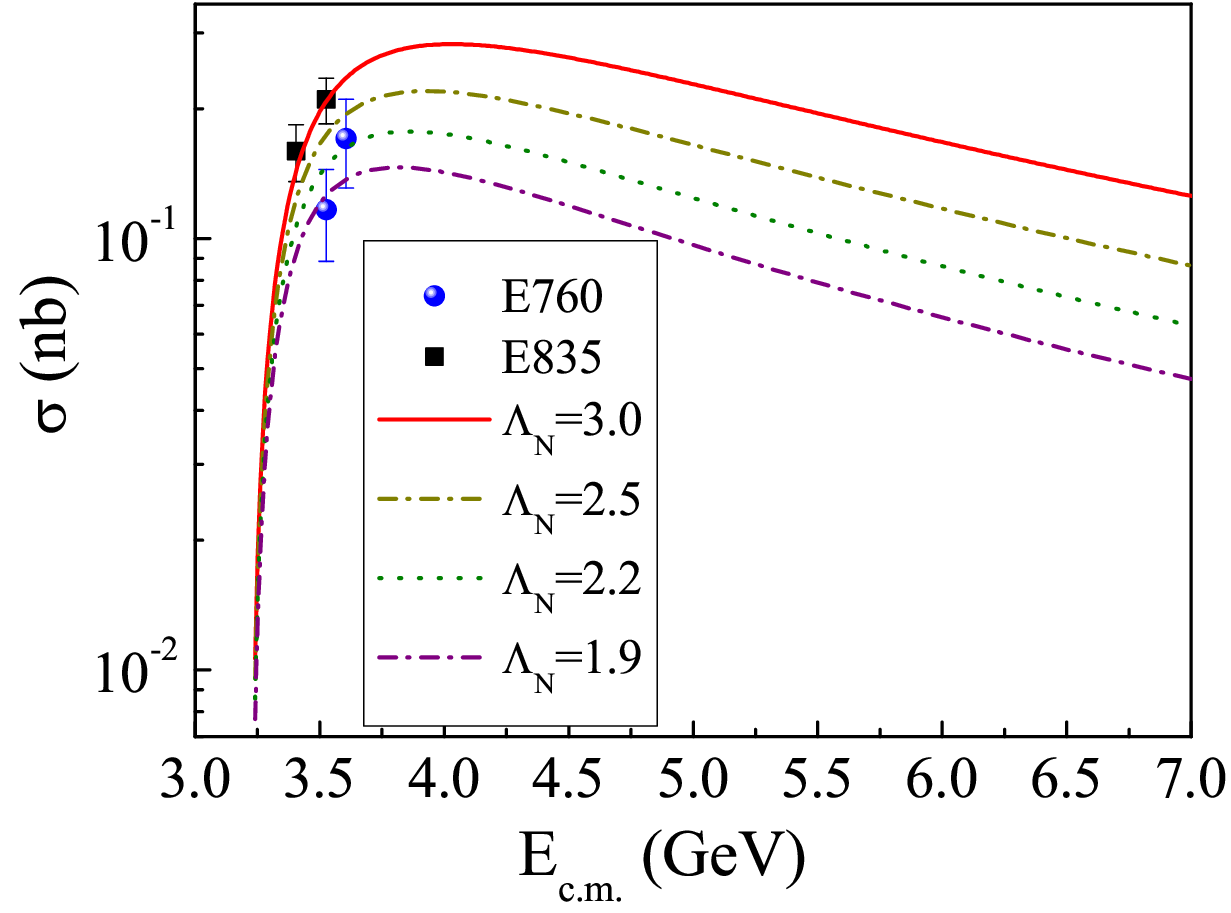}
\caption{(Color online) The contribution of background from the proton
exchange for $\bar{p}p\rightarrow J/\protect\psi \protect\pi ^{0}$ reaction
though t-channel and u-channel with different typical cut-off $\Lambda _{N}$%
. The datas are taken from Ref. \protect\cite{E760,E835}}
\end{figure}

\begin{figure}[tbph]
\centering
\includegraphics[scale=0.4]{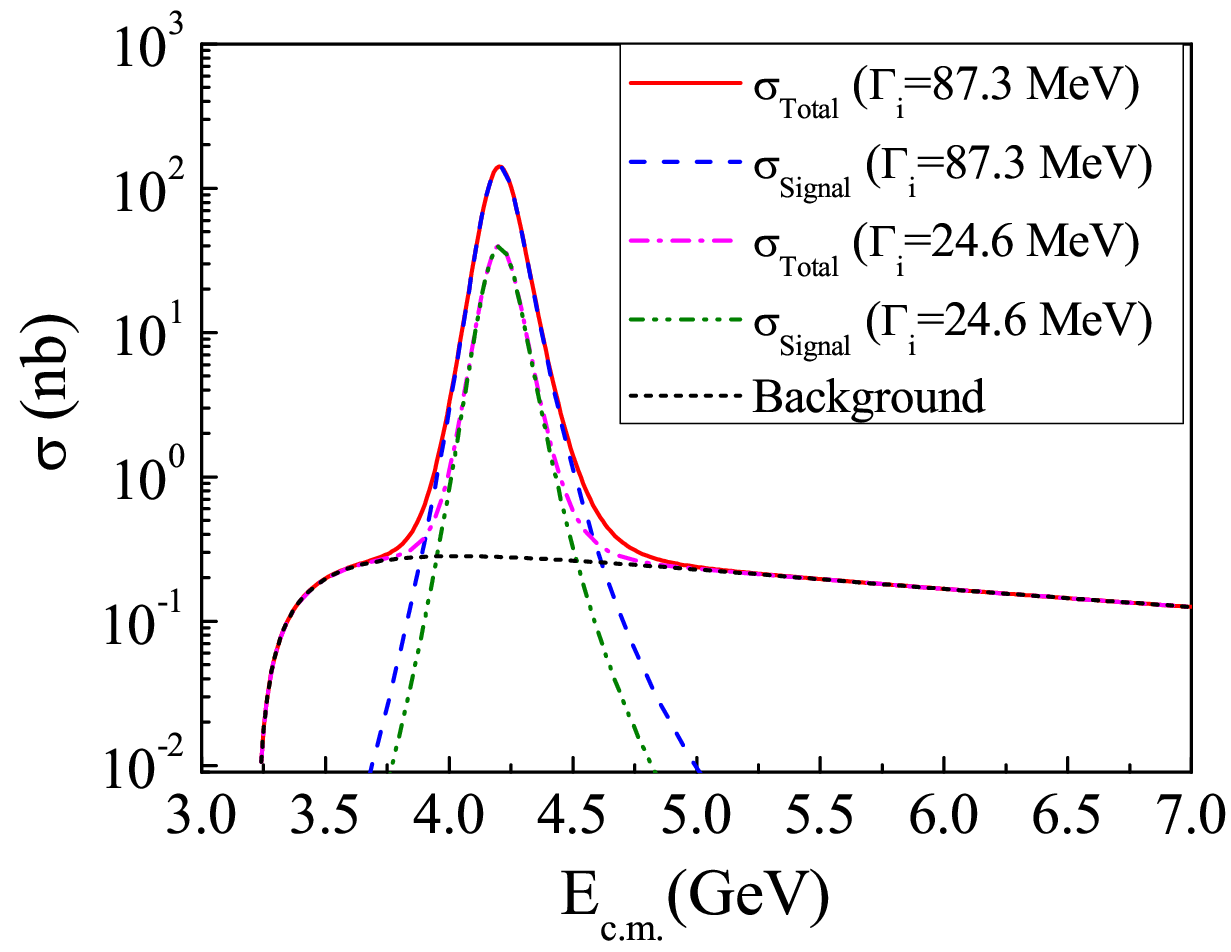}
\caption{(Color online) The energy dependence of the total cross sections
for the process of $\bar{p}p\rightarrow J/\protect\psi \protect\pi ^{0}$
with two typical values of $\Gamma _{i}$, where $\Gamma _{i}$ denotes the
decay width of $Z(4200)\rightarrow J/\protect\psi $. Here, the $\protect%
\sigma _{Total}$ and $\protect\sigma _{\text{S}ignal}$ are the total cross
section and the cross section of signal, respectively.}
\end{figure}

With the formalisms and equations determined above, we calculate the total
and differential cross section including both signal and background
contributions as presented in Figs 4-7.

In these calculations, we note that the cut-off parameter related to the
form factor is the only free parameter. Therefore, first we need to discuss
the effect of cut-off parameter on cross section of signal and background.

We present the variation of the cross section from the s-channel signal
contribution for $\bar{p}p\rightarrow J/\psi \pi ^{0}$ reaction with
different cut-off parameters $\Lambda _{Z}$ as shown in Fig. 4, where $%
\Lambda _{Z}$ is taken as 1.0-3.0 GeV with the step of 1.0 GeV.

One notice that there is a obvious peak structure at center-mass energy $%
E_{c.m.}\simeq 4.2$ GeV which is near the threshold of $Z(4200)$. Moreover,
the cross section of signal increases with the increasing of cut-off
parameter $\Lambda _{Z},$ but at a modest rate. Especially in the range of $%
4.0$ $GeV\lesssim E_{c.m.}\lesssim 4.5$ $GeV$, it is found that the cross
sections from signal contributions are not sensitive to the cut-off
parameter $\Lambda _{Z}$. We take typical value $\Lambda _{Z}=1.0$ GeV in
the following, which can ensure the cross section of signal are limited to a
smaller value.

In Fig. 5, we illustrate the proton exchange contributions with different
cut-off parameters $\Lambda _{N}$, which is obvious that the cross sections
from background contributions are sensitive to the values of the cutoff $%
\Lambda _{N}$. Fortunately, the reaction $\bar{p}p\rightarrow J/\psi \pi
^{0} $ has been measured by the E760 and E835 experiment at low energy \cite%
{E760,E835}, which can help us to constrain the cut-off parameter $\Lambda
_{N}$. From Fig. 5 it can be found that the numerical results from the
proton exchange contributions are consistent with the E760 and E835 datas by
taking $\Lambda _{N}=1.9$ and $3.0$ GeV, respectively. Based on the
consideration of seeking a larger limit for the cross section of background,
we take $\Lambda _{N}=3.0$ GeV in the next calculation. Besides, we notice
that the amplitude estimate of $\bar{p}p\rightarrow J/\psi \pi ^{0}$ is
about 0.3 $nb$ at $E_{cm}=3.5-3.6$ GeV in ref. \cite{al06}. This value is
closer to the E835 data \cite{E835} if we consider its uncertainty. Thus we
taking $\Lambda _{N}=3.0$ GeV in our calculation should be reasonable.

Fig. 6 show the total cross sections for $\bar{p}p\rightarrow J/\psi \pi
^{0} $ reaction including both signal and background contributions by taking
$\Lambda _{Z}=1.0$ GeV and $\Lambda _{N}=3.0$ GeV. We notice that the cross
section of $Z^{0}(4200)$ production goes up very rapidly and has a peak
around $E_{cm}\simeq 4.2$ GeV. Besides, it is found that the contributions
from the signal are dominant in the region of $4.0$ GeV $\lesssim
E_{c.m.}\lesssim 4.5$ GeV. Naturally, we can conclude that $4.0$ GeV $%
\lesssim E_{cm}\lesssim 4.5$ GeV is the best energy window for searching the
neutral charmoniumlike $Z^{0}(4200)$ in experiment, which the signal can be
clearly distinguished from background. Around the center of mass $%
E_{cm}\simeq 4.2$ GeV, the total cross section from signal and background
contributions is on the order of 0.14 $\mu b$ and 0.04 $\mu b$, which
correspond to the decay width $\Gamma _{Z^{0}(44200)\rightarrow J/\psi \pi
^{0}}=87.3$ MeV and $\Gamma _{Z^{0}(44200)\rightarrow J/\psi \pi ^{0}}=24.6$
MeV, respectively.

As mentioned above, the PANDA detector at FAIR \cite{panda,uw11,cs12,gy08}
is an ideal platform searching for the $Z^{0}(4200)$ by $\bar{p}p$
collision. With a $\bar{p}$ beam of 15 GeV$/c$ \cite{panda,uw11,cs12,gy08}
one has $E_{c.m.}=5.47$ GeV, which allows one to observe charmoniumlike $%
Z^{0}(4200)$ state in $J/\psi \pi ^{0}$ production up to a mass $M_{Z}\simeq
4.2$ GeV. Assuming the integrated luminosity of PANDA can reach up to 1.5 fb$%
^{-1}$ per year \cite{panda,uw11,cs12,gy08}, taking $\sigma _{total}\approx
0.04-0.14$ $\mu b$, one can expect about $6\times 10^{7}-2.1\times 10^{8}$
events per year for the production of $J/\psi \pi ^{0}$ at $E_{c.m.}\simeq
4.2$ GeV, which are enough to meet the requirement of the experiment. At
present, except for the $Z(4200)$, other charmonium-like Z states (such as $%
Z(3900)$, $Z(4025)$ and $Z(4430)$ etc.) were also observed in the $J/\psi
\pi $ invariant mass. Since all of them have probably the identical quantum
numbers $J^{P}=1^{+}$, the interference of each other is possible. The above
situation indicate that the contributions from other $Z^{0}$ maybe be
significant for the $\bar{p}p\rightarrow J/\psi \pi ^{0}$ process. However,
in this work, we focus only on the $Z^{0}(4200)$ state and do not consider
the contributions from other $Z^{0}$ states.
\begin{figure*}[tbph]
\begin{minipage}{1\textwidth}
\includegraphics[scale=0.38]{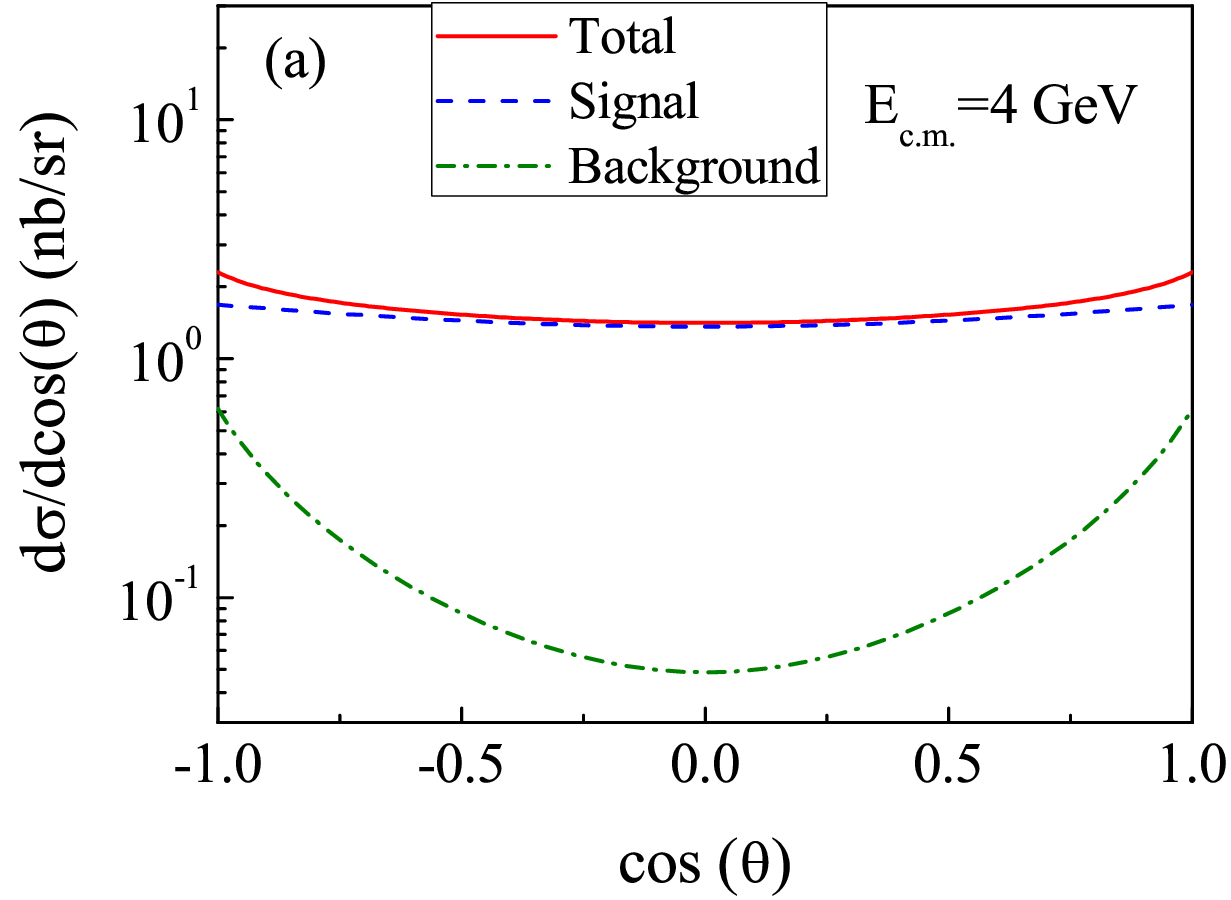}
\includegraphics[scale=0.38]{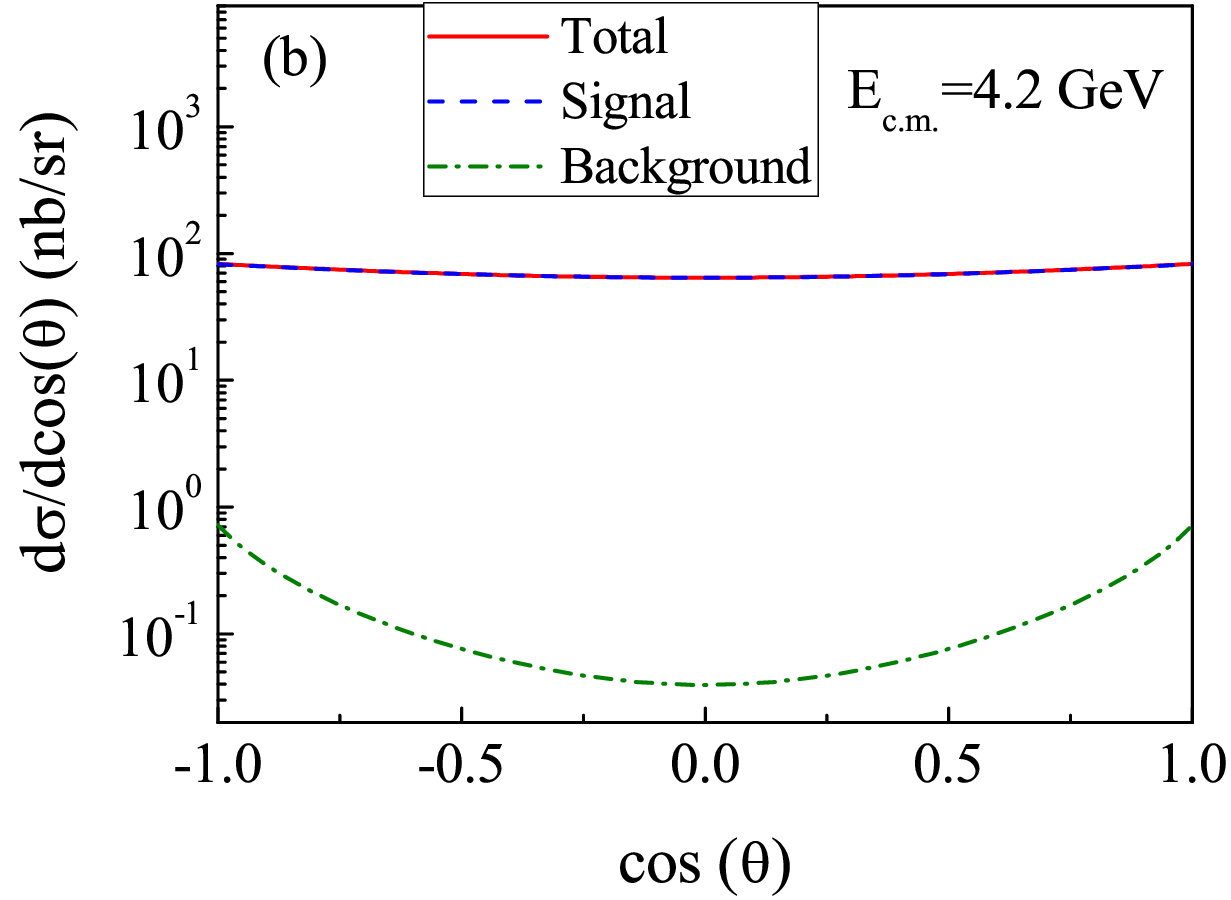}
\includegraphics[scale=0.38]{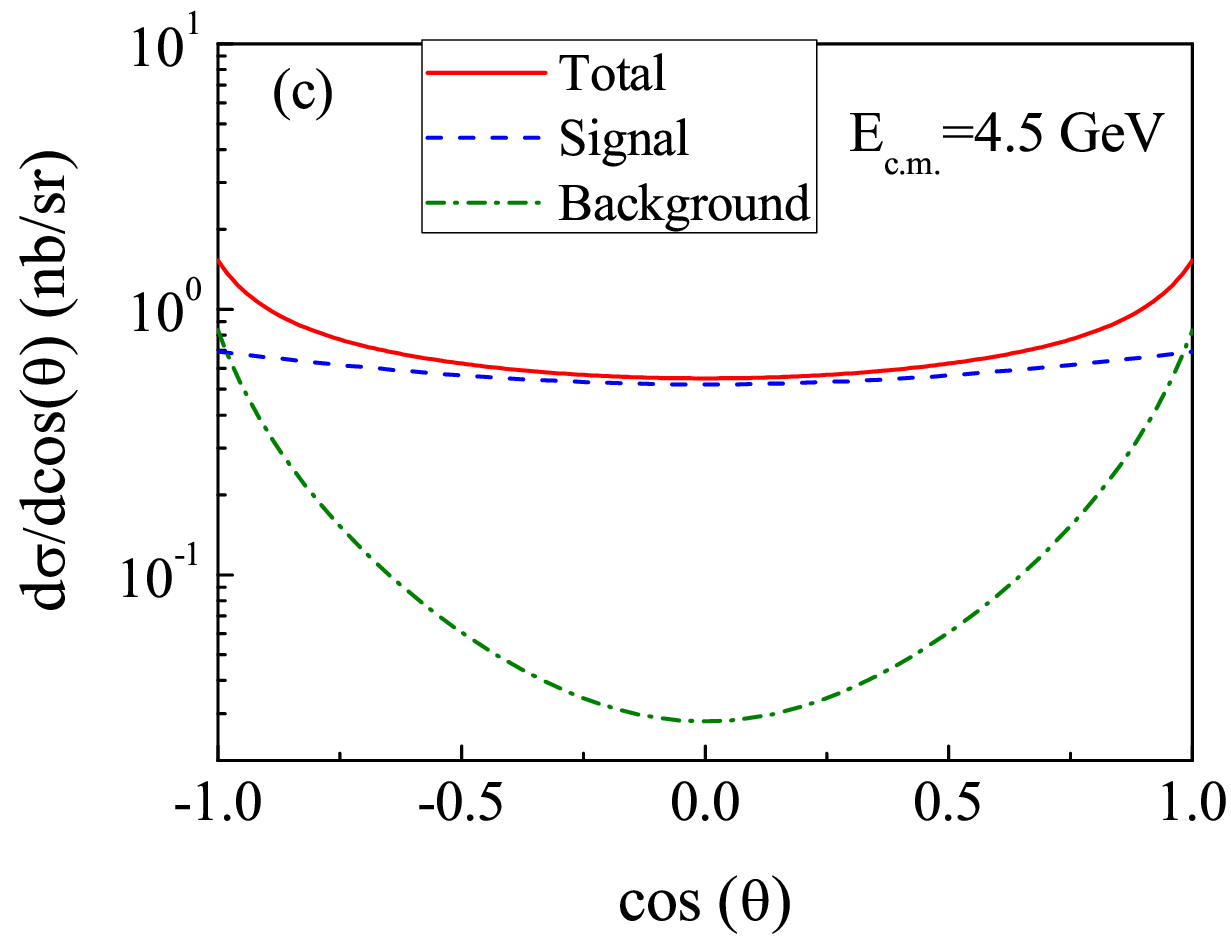}
\includegraphics[scale=0.38]{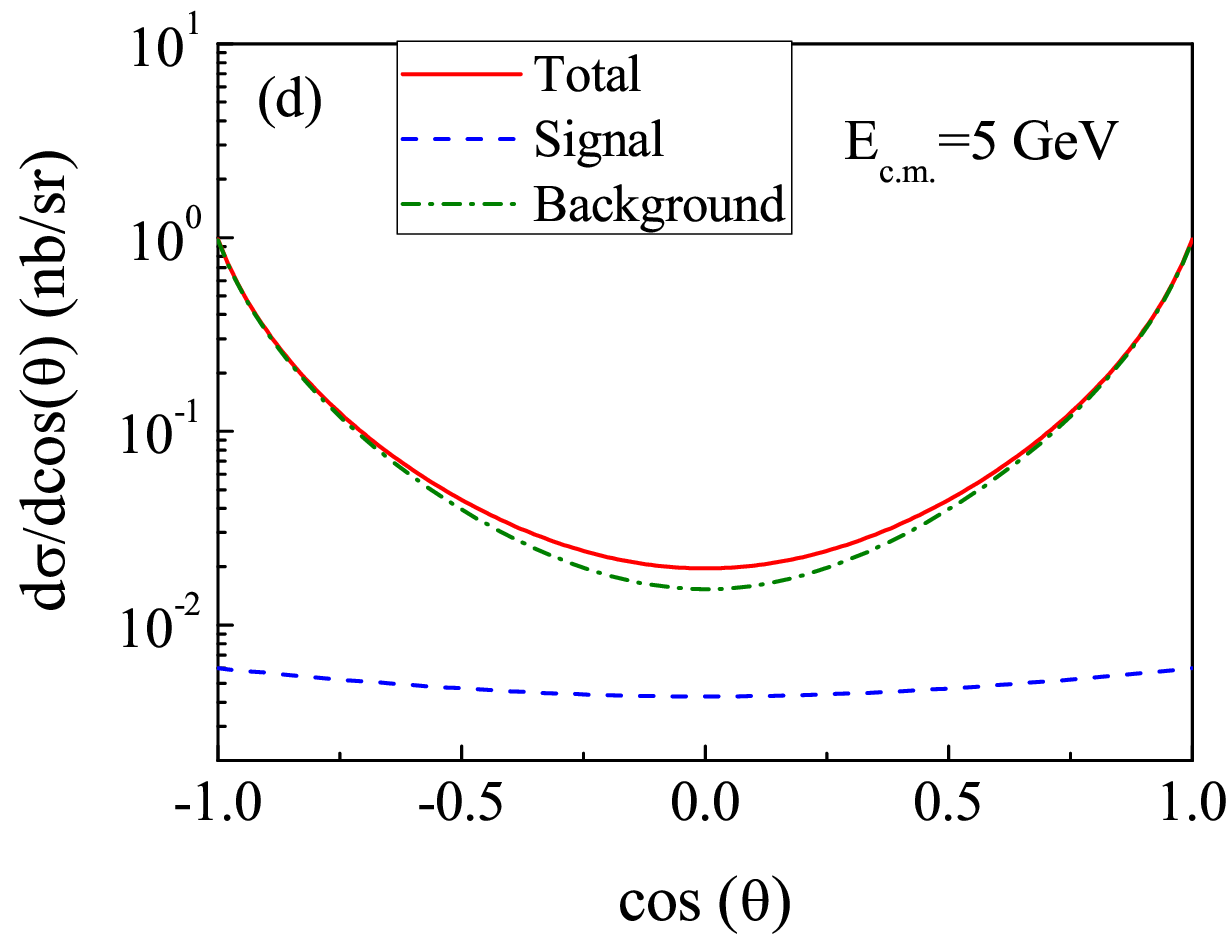}
\caption{(Color online) The differential cross sections for the process of $%
\bar{p}p\rightarrow \protect J\psi\protect\pi ^{0}$ at different
center of mass energy $E_{c.m.}=4.0$, $4.2$, $4.5$ and $5$ GeV, where the "Total"
denotes the differential cross section including both signal and background
contributions. Here, the partial decay width is taken as $\Gamma
_{Z(4200)\rightarrow J/\protect\psi \protect\pi }=87.3$ MeV.}
\end{minipage}
\end{figure*}

Fig. 7 show the differential cross section including both signal and
background contributions at the center of mass energy $E_{c.m.}=4.0$, 4.2,
4.5 and 5 GeV. We notice that the line shape of total differential cross
section are less affect by background and almost coincident with the line
shape of signal differential cross section at $E_{c.m.}=4.0-4.5$ GeV, which
are consistent with the calculations as presented in Fig. 6. In comparison,
it is found that the shapes of total angular distributions is different from
the shapes of signal angular distributions at $E_{c.m.}=5$ GeV$,$ which due
to the background has a strong effect on the total cross section at $%
E_{c.m.}=5$ GeV. These predictions can be checked by the future experiment.
\

\section{Discussion and Conclusion}

In this work, we investigate the neutral $Z^{0}(4200)$ production in $\bar{p}%
p\rightarrow J/\psi \pi ^{0}$ reaction with an effective Lagrangian
approach. Our numerical result indicate that the $\bar{p}p\rightarrow J/\psi
\pi ^{0}$ is very likely an ideal channel to study and search for the
neutral hidden charm $Z^{0}(4200)$. Furthermore, it is found that the center
of mass energy $E_{c.m.}\simeq 4.0-4.5$ GeV is the best energy window for
searching the neutral $Z^{0}(4200)$, which the signal can be easily
distinguished from background. Moreover, According to our estimation, enough
$Z^{0}(4200)$ events near $E_{c.m.}\simeq 4.2$ GeV can be produced at PANDA,
which indicate that searching for the neutral $Z^{0}(4200)$ by $\bar{p}p$
annihilation at PANDA is feasible. Besides, since our calculations are
carried out in the premise of assuming that the $Z^{0}(4200)$ has a coupling
with $\bar{p}p$ and $J/\psi \pi ^{0}$, the near future experiments at LHC
and BelleII will be able to check our predictions on the respective coupling
strengths of the $Z^{0}(4200)$.

It should be mentioned that the value of coupling constant $g_{Z\bar{p}p}$
is determined by analyzing and comparing the degree of OZI suppressed in the
process of $Z^{0}(4200)\rightarrow \bar{p}p$ and $J/\psi \rightarrow \bar{p}%
p $, which is based on the assumption of the neutral $Z^{0}(4200)$ may be
composed by $\left\vert c\bar{c}u\bar{u}\right\rangle $ or $\left\vert c\bar{%
c}d\bar{d}\right\rangle $. According our estimate, even if taking $\Gamma
_{Z^{0}(4200)\rightarrow \bar{p}p}=20$ keV (this value is one order smaller
than that we used in the above calculations ) and $\Gamma
_{Z^{0}(44200)\rightarrow J/\psi \pi ^{0}}=24.6$ MeV, the cross section from
signal $Z^{0}(4200)$ contributions by $\bar{p}p$ annihilation at $%
E_{c.m.}\simeq 4.2$ GeV could still reach up to the level of 4 nb which is
at least one order higher than that of background. This means that an
obvious bump at $E_{c.m.}\simeq 4.2$ GeV can be expected to be appear in the
$\bar{p}p\rightarrow J/\psi \pi ^{0}$ process. Thus, we strong suggest the
relevant experiment can be carried out, which will not only be conducive to
verify the existence of $Z(4200)$ but enable us to have a more comprehensive
understanding of the nature of exotic states and OZI rule.

\section{Acknowledgments}

The authors would like to acknowledge Ju-Jun Xie for useful comments. The
author X. Y. Wang is grateful Dr.Qing-yong Lin for valuable discussions and
help. This project is partially supported by the National Basic Research
Program (973 Program Grant No. 2014CB845406), the National Natural Science
Foundation of China (Grants No. 11175220) and the the one Hundred Person
Project of Chinese Academy of Science (Grant No. Y101020BR0).

\end{document}